\documentclass{iopart}

\usepackage{graphicx}
\usepackage{subfig}

\usepackage{amsopn}
\usepackage{epsfig}
\usepackage{amsmath} 
\usepackage{breqn}
\usepackage{epstopdf}
\usepackage{bm}
\usepackage{amssymb}
\usepackage{color}
\usepackage{cite}
\usepackage{enumitem}

\usepackage{makecell}
\usepackage{geometry}
\usepackage[abbr]{harvard}
\citationstyle{agsm}
\renewcommand{\harvardand}{and}

\usepackage{algorithm,algorithmic}
\usepackage{epsfig,times,caption}
\usepackage{algorithm,algorithmic}
\usepackage{amsfonts}
\usepackage{booktabs}
\usepackage{tabularx}

\begin{document}
\title{Head and Neck Tumor Segmentation from [$^{18}$F]F-FDG PET/CT Images Based on 3D Diffusion Model}
\author{Yafei Dong$^{1}$, Kuang Gong$^{2,3}$}
\address{$^1$Yale PET Center, Department of Radiology and Biomedical Imaging, Yale University School of Medicine, New Haven, CT, 06520, USA}
\address{$^2$J. Crayton Pruitt Family Department of Biomedical Engineering, University of Florida,  Gainesville, FL, 32611, USA \\}%
\address{$^3$Gordon Center for Medical Imaging, Department of Radiology, Massachusetts General Hospital, Boston, MA, 02114, USA}%
\ead{kgong@bme.ufl.edu}

\begin{abstract}

\textit{Objective.} Head and neck (H\&N) cancers are among the most prevalent types of cancer worldwide, and [$^{18}$F]F-FDG PET/CT is widely used for H\&N cancer management. Recently, the diffusion model has demonstrated remarkable performance in various image-generation tasks. In this work, we proposed a 3D diffusion model to accurately perform H\&N tumor segmentation from 3D PET and CT volumes. 
\textit{Approach.} The 3D diffusion model was developed considering the 3D nature of PET and CT images acquired. During the reverse process, the model utilized a 3D UNet structure and took the concatenation of 3D PET, CT, and Gaussian noise volumes as the network input to generate the tumor mask. Experiments based on the HECKTOR challenge dataset were conducted to evaluate the effectiveness of the proposed diffusion model. Several state-of-the-art techniques based on  U-Net and Transformer structures were adopted as the reference methods. Benefits of employing both PET and CT as the network input, as well as further extending the diffusion model from 2D to 3D, were investigated based on various quantitative metrics and qualitative results.
\textit{Main results.} Results showed that the proposed 3D diffusion model could generate more accurate segmentation results compared with other methods (mean Dice of 0.739 compared to less than 0.726 for other methods). Compared to the diffusion model in 2D form, the proposed 3D model yielded superior results (mean Dice of 0.739 compared to 0.669). Our experiments also highlighted the advantage of utilizing dual-modality PET and CT data over only single-modality data for H\&N  tumor segmentation (with mean Dice less than 0.570).
\textit{Significance.} This work demonstrated the effectiveness of the proposed 3D diffusion model in generating more accurate H\&N tumor segmentation masks compared to the other reference methods.
\end{abstract}

\section{Introduction}
\label{sec:intro}  
Head and neck (H\&N) cancer is a comprehensive term encompassing malignancies that originate in the paranasal sinuses, nasal cavity, oral cavity, pharynx, and larynx. It ranks as the seventh most common type of cancer worldwide~\cite{mody2021head}. In 2020 alone, it was estimated that approximately 65630 new cases of oral cavity, pharyngeal, and laryngeal cancers would occur, accounting for about 3.6\% of all new cancer cases in the United States~\cite{pfister2020head}. [$^{18}$F]F-FDG PET and CT imaging stand out as the widely used modalities for the initial staging and subsequent monitoring of H\&N cancer \cite{gupta2011diagnostic}.  [$^{18}$F]F-FDG PET images can effectively highlight H\&N tumors with high metabolic activities, but suffer from low image resolution and signal-to-noise ratio (SNR).  Occasionally,  distinguishing H\&N tumors from other highly metabolically active normal tissues can be challenging based on [$^{18}$F]F-FDG PET images. On the other hand, CT imaging offers valuable morphological insights into the structures of the human body. Consequently, integrating both  [$^{18}$F]F-FDG PET and CT images can synergistically provide complementary information, enhancing the accuracy of diagnosis and treatment planning. Accurate identification of a mass lesion and estimation of tumor volume significantly impact subsequent treatment planning. Inaccurate positionin of tumors can lead to substantial underdosing of the tumor or overdosing of critical organs during radiation treatment. However, it is a laborious job for radiologists to delineate the tumor volume from PET and CT images for every patient, especially when dealing with large cohorts. Hence, the development of highly accurate automatic segmentation techniques for H\&N tumors holds great significance, which can facilitate faster and more reproducible tumor delineation.

Recently, deep learning methods built on convolutional neural networks (CNNs) have achieved impressive performance for medical image segmentation tasks. Ronneberger et al.~\cite{ronneberger2015u} proposed the U-Net model for the segmentation of neuronal structures in electron microscopic stacks. It consists of an encoder,  a decoder, and skip connections from the encoder to the decoder to restore the lost information. Over the years, the U-Net model has become the most widely adopted image segmentation architecture. Various extensions of U-Net have been proposed to handle the domain-specific complexity inherent in different segmentation tasks~\cite{azad2022medical,guo2019gross,zhou2019unet++,huang2020unet,xiang2020bio,oktay2018attention,li2020attention,cciccek20163d}. The processing steps and parameter settings of U-Net models heavily relies on researchers' expertise, particularly in cases where substantial dataset disparities exist. As a solution to this challenge, Isensee et al.~\cite{isensee2021nnu} proposed a self-configuring U-Net model named nnU-Net, which has the capability to automatically configure itself regarding preprocessing, network architecture, training, and post-processing. 

One potential issue of CNN-based segmentation networks is the limited receptive field through convolution operations. The Transformer networks, initially proposed for natural language processing (NLP)~\cite{vaswani2017attention}, have a larger receptive field and can better utilize long-range spatial information than CNN. Researchers have extended Transformer networks to various computer vision tasks, including medical image denoising and segmentation~\cite{jang2023spach,chen2021transunet,dosovitskiy2020image,hatamizadeh2022unetr,hatamizadeh2021swin}. Chen et al.~\cite{chen2021transunet} proposed the Trans\-UNet model, which integrated the stacked vision transformer (ViT)~\cite{dosovitskiy2020image} within the middle part of U-Net to address U-Net's limitation in capturing long-range information. For 3D medical image segmentation, Hatamizadeh et al.~\cite{hatamizadeh2022unetr} introduced a model named UNEt TRansformers (UNETR). This model adopted a UNet-like structure, utilizing ViT as the encoder to capture global information within 3D volumes. The Swin UNETR~\cite{hatamizadeh2021swin} is a modification to the original UNETR, where the 3D ViT was replaced by the Swin transformer~\cite{liu2021swin}. The Swin transformer encoder can effectively extract features at different distinct resolutions by shift-window operations during self-attention calculation.

The diffusion model is one category of generative models and was initially proposed for image synthesis~\cite{ho2020denoising}. It decomposed a challenging image synthesis task into a sequence of image denoising subtasks, leading to more stable convergence compared to some conventional methods, reducing the risk of model collapse and overfitting. Therefore, the diffusion model showed competitive performance when compared with other state-of-the-art methods~\cite{dhariwal2021diffusion} in the image synthesis task. Additionally, because of its generative nature, diffusion models can produce multiple outputs from the same input, which can be beneficial in situations where uncertainty or ambiguity is inherent to the task. This is in contrast to CNNs and Transformer networks, which are deterministic and yield only a single result for a given input. Beyond image synthesis, they also demonstrated significant potentials across various domains, e.g., image super-resolution~\cite{saharia2022image}, denoising~\cite{gong2023pet},  inpainting~\cite{rombach2022high} and segmentation~\cite{wu2022medsegdiff,li2024generic,zhao2024dtan,khosravi2023few}. For all the aforementioned diffusion models-based approaches, the processing primarily focused on 2D slices in the transverse plane. This approach ignored the information existing between different axial slices, which is highly beneficial for 3D medical segmentation tasks.

In this work, we extended the diffusion model to the 3D form for H\&N tumor segmentation, where 3D convolutional layers were deployed to extract features from three dimensions simultaneously based on both 3D [$^{18}$F]F-FDG PET and CT images. 
To assess performance of the proposed 3D diffusion model, experiments were conducted using a publicly available H\&N clinical dataset. Regarding reference methods, the proposed 3D diffusion model was compared with state-of-the-art U-Net- and Transformer-based approaches, including U-Net, nnU-Net, UNETR, Swin UNETR, and Pengy U-Net. Benefits of 3D over 2D for diffusion models regarding performance and uncertainty evaluation were also investigated. Finally, benefits of supplying PET and CT images together as the input of diffusion models were also evaluated with that of utilizing PET or CT images alone.

\section{Method}
\subsection{Diffusion Models}

The denoising diffusion probabilistic model (DDPM)~\cite{ho2020denoising} is one popular framework of diffusion models, which was adopted in our work. DDPM consists of two processes: a forward process that gradually adds Gaussian noise to the clean data (such as signals, images, or volumes) via a predefined Markov chain and a backward process that tries to iteratively recover the original data from noise. 
During the forward process, given the clean data $x_0$ sampled from the distribution $q(x_0)$, small amounts of Gaussian noise are gradually added to it in $T$ timesteps using a predefined Markov chain as:
\begin{align}
q(x_{1:T} | x_0) &= \prod_{t=1}^{T} q(x_t | x_{t-1}), \\
q(x_t | x_{t-1}) &= \mathcal{N} (x_t ; \sqrt{1-\beta_t}x_{t-1}, \beta_t \mathbf{I}),
\end{align}
where $\mathcal{N}(x;\mu,\sigma^2)$ denotes the Gaussian distribution $x$ with mean $\mu$ and variance $\sigma^2$, and $\beta_t$ is a hyper-parameter to control the variance of the Gaussian noise. $\mathbf{I}$ denotes the unit matrix. When $T$ becomes large enough, the distribution of $x_T$ can be regarded as an isotropic Gaussian distribution.
 With $\alpha_t = 1 - \beta_t$ and $\overline{\alpha}_t = \prod_{i=1}^t \alpha_i$, the forward process can be rewritten as:
\begin{equation}
q(x_t | x_0) = \mathcal{N} (x_t; \sqrt{\overline{\alpha}_t}x_0, (1-\overline{\alpha}_t) \mathbf{I}).
\end{equation} 
{\color{black} During the reverse process, a neural network with learnable parameters $\theta$ is trained to recover the clean data $x_0$ from the corrupted data $x_T$ (sampled from the Gaussian distribution) step by step.}
To constrain the stochasticity from random-sampled $x_T$, the conditional data $y_t$ is provided as the input of the neural network together with $x_t$ and the time step $t$. Therefore, the reverse process is defined as:
\begin{align}
p_\theta (x_{0:T}) &= p(x_{T}) \prod_{t=1}^{T} p_\theta (x_{t-1} | x_t, y_t), \\
p(x_{T}) &= \mathcal{N}(x_{T};\mathbf{0},\mathbf{I}), \\
p_\theta (x_{t-1} | x_t, y_t) &= \mathcal{N} (x_{t-1};\mu_{\theta} (x_t,y_t,t), \Sigma_{\theta} (x_t,y_t,t)),
\end{align}
where $\mu_{\theta} (x_t,y_t,t)$ and $\Sigma_{\theta} (x_t,y_t,t))$ denote the predicted mean and variance value by the neural network respectively from input $x_t$, $y_t$, and $t$. Nevertheless, in practice, we tend to train the model to predict the added noise instead of mean and variance, and the parameters $\theta$ of the neural network is trained by minimizing the difference between the added noise $\epsilon_t$ and the estimated noise $\epsilon_{\theta} (x_t,y_t,t)$ in every step $t$ as follows:
\begin{equation}
\mathcal{L}(\theta) = \sum_t\| \epsilon_t - \epsilon_{\theta} (x_t,y_t,t) \|^2 .
\end{equation}
After getting the $\epsilon_{\theta} (x_t,y_t,t)$, the mean value $\mu_{\theta} (x_t,y_t,t)$ can be derived as:
\begin{equation}
\mu_{\theta} (x_t,y_t,t) = \frac{1}{\sqrt{\alpha_t}}(x_t - \frac{1-\alpha_t}{\sqrt{1-\overline{\alpha}_t}}\epsilon_{\theta} (x_t,y_t,t)) .
\end{equation}
Besides, in~\cite{ho2020denoising}, Ho et al. chose to manually set $\Sigma_{\theta} (x_t,y_t,t))$ as $\beta_t \mathbf{I}$. Therefore, finally, the $x_{t-1}$ can be calculated as:
\begin{equation}
x_{t-1} = \frac{1}{\sqrt{\alpha_t}}(x_t - \frac{1-\alpha_t}{\sqrt{1-\overline{\alpha}_t}}\epsilon_{\theta} (x_t,y_t,t)) + \sqrt{\beta_t} \mathbf{z},
\end{equation}
where $\mathbf{z} \sim \mathcal{N}(\mathbf{0},\mathbf{I})$. By repeating the aforementioned process $T$ times, the clean data $x_0$ can be restored from noise.
\begin{figure*}
	\centering
	\includegraphics[width=.99\textwidth]{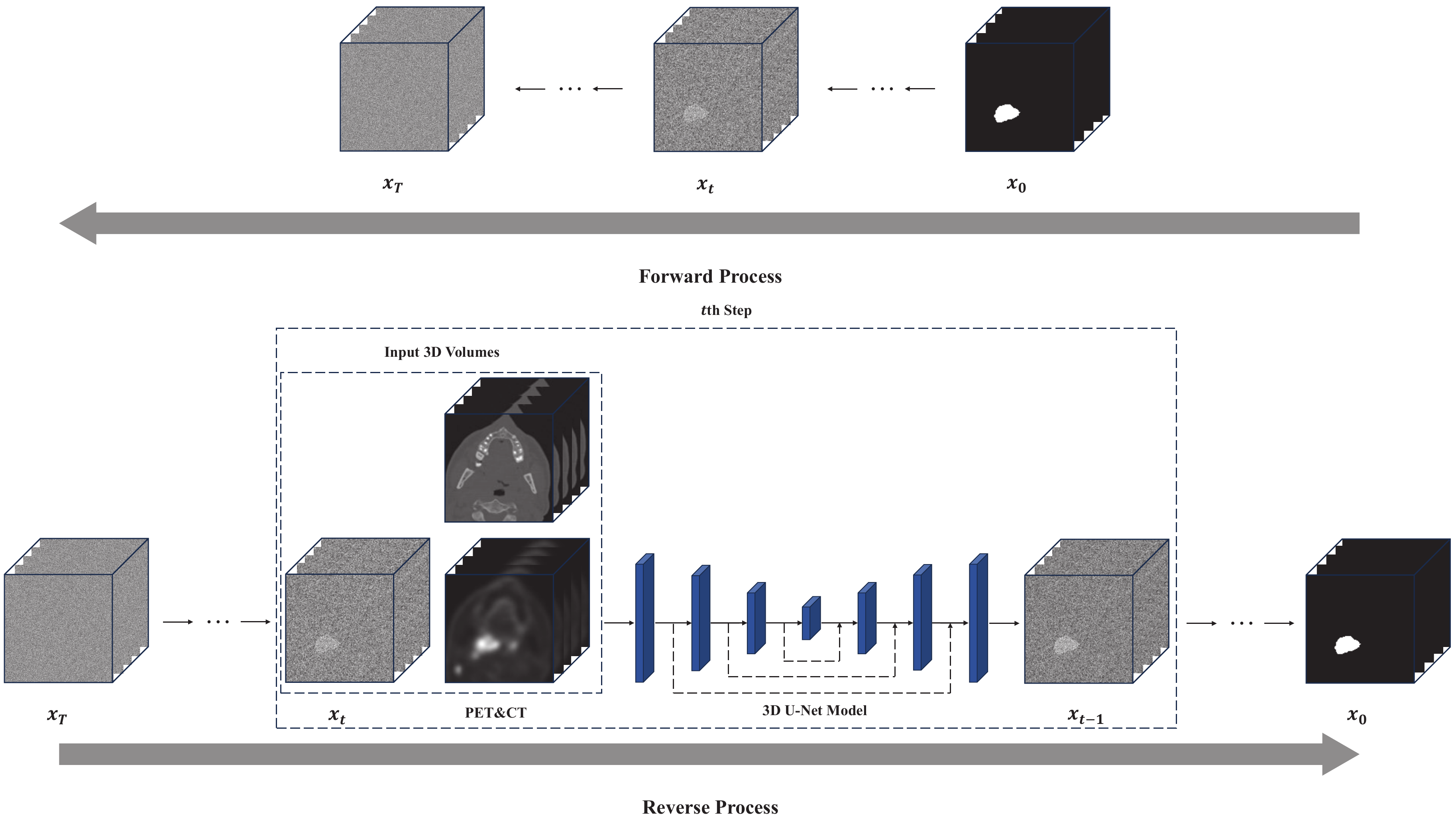}
	\caption {Graphic illustration for our proposed 3D diffusion model. In the forward process, small amounts of Gaussian noise were added to the 3D tumor masks. In the reverse process, the model took 3D [$^{18}$F]F-FDG-PET, CT, and Gaussian distribution volumes as input and restored the 3D tumor masks step by step.}
	\label{fig_ddpm}
\end{figure*} 

\begin{table}[t]
\centering
\caption{List of detailed hyper-parameters of the proposed 3D diffusion model. For input and output dimensions, $N$ denotes the batch size, and $D$, $H$, and $W$ represent dimensions in the depth, height, and width directions, respectively. The base channels represent the dimensions of the initial layer, whereas the channel multipliers serve as factors that are multiplied by the base channels for subsequent layers.}
\begin{tabular}{@{}lc@{}}
\toprule
Hyper-parameter           & Value \\ \midrule
Input Dimension           & $N \times 3 \times D \times H \times W$     \\
Output Dimension          & $N \times 1 \times D \times H \times W$     \\
Convolutional Kernel Size & $(3,3,3)$     \\
Down/up-sample Kernel Size       & $(1,2,2)$     \\
Diffusion Steps           & 1000     \\
Noise Schedule            & Linear     \\
Base Channels             & 128     \\
Channel Multiplier        & $(1, 1, 2, 3, 4)$     \\
Residual Blocks           & 2     \\
Model Size                & 214M     \\ \bottomrule
\end{tabular}
\label{tab_model}
\end{table}

\subsection{Network Structure}

The graphic illustration of the proposed 3D diffusion model is shown in Figure~\ref{fig_ddpm}. For our H\&N segmentation task, both input and output data were represented in 3D form with size $D\times H \times W$, where $D$, $H$, and $W$ corresponded to dimensions in the depth (i.e., number of axial slices), height, and width directions, respectively. The targeted clean data $x_0$ was the 3D tumor mask. 
{\color{black} During training, in the forward process, we gradually introduced small amounts of Gaussian noise to the 3D tumor mask $x_0$, resulting in the distorted data $x_T$ as illustrated at the top of Figure~\ref{fig_ddpm}. In the reverse process, a neural network with parameters $\theta$ was employed to reconstruct the clean data $x_0$ from $x_T$. The 3D volumes of [$^{18}$F]F-FDG PET, CT, and $x_t$ were concatenated in the channel dimension as the input for the neural network, with a size of $3\times D\times H \times W$, as described at the bottom of Figure~\ref{fig_ddpm}. During inference, the reverse process started from a randomly sampled Gaussian distribution with a mean of 0 and a standard deviation of 1 for $x_T$, generating the clean segmentation $x_0$ step by step.}
Typically, a 2D U-Net model is utilized in the reverse process. For our specific 3D segmentation task, we applied a 3D U-Net model to effectively capture all information across the three dimensions. The convolutional layers in this model employed 3D operations with kernel sizes of $(3, 3, 3)$. The kernel sizes for the downsample and upsample operations were set to $(1, 2, 2)$, trying to preserving the information along the axial direction. Additionally, we adopted the residual block as the fundamental unit in the model, as introduced in~\cite{brock2018large}. Each block included group normalization layers~\cite{wu2018group}, SiLU activation layers~\cite{elfwing2018sigmoid}, convolutional layers, and the residual skip connection.
Regarding the hyper-parameters during the forward and reverse process, we designated a total of 1000 time steps ($T = 1000$) and implemented a linear noise schedule for $\beta_t$. Table~\ref{tab_model} shows a detailed list of parameter settings of our model.

\subsection{Dataset and Prepocessing}

The HEad and neCK TumOR 2021 (HECKTOR 2021) challenge dataset~\cite{andrearczyk2021overview} was utilized in the experiment. This dataset comprised of 224 cases with H\&N tumor located in the oropharynx region sourced from five centers at Canada and France (HGJ, CHUS, HMR, CHUM and CHUP). Each case included the co-registered 3D [$^{18}$F]F-FDG PET volume and 3D CT volume focusing on the head and neck region as well as a binary contour outlining the annotated ground truth of the tumor. The [$^{18}$F]F-FDG PET volumes were expressed in Standardized Uptake Values (SUV), while the CT volumes were measured in Hounsfield Units (HU). A bounding box pinpointing the oropharynx region was also included. Besides, patient clinical data were also provided for all cases including center, age, gender, TNM 7/8th edition staging and clinical stage, tobacco and alcohol consumption, performance status, HPV status, and treatment (radiotherapy only or chemoradiotherapy). According to the clinical data, there were 167 male cases and 57 female cases among all 224 cases, and the average age of the patients in the dataset was 62.7 years old. Out of these cases, 84 had a positive HPV status. The segmentation task poses challenges, including variations in image acquisition and quality across centers, as well as the presence of lymph nodes exhibiting high metabolic responses in the PET images. Across the five centers, PET/CT scanners from GE Healthcare were used by three centers, while one center utilized a scanner from Philips and another utilized a scanner from Siemens. To evaluate various models' performances, we manually divided the dataset into training, validation, and testing sets of 170, 14, and 40 cases, respectively. We resampled the data to the isotropic resolution of $1\times1\times1~\mbox{mm}^3$ and the size of $144\times144\times144$ by using the trilinear interpolation. The values of CT and PET data were normalized to the range of [-1,1]. 

\subsection{Reference Methods}

We conducted a comparative analysis between our proposed 3D DDPM method and other state-of-the-art U-Net- and Transformer-based methods in the field of medical image segmentation. The reference models included the plain 3D U-Net~\cite{ronneberger2015u}, 3D nnU-Net~\cite{isensee2021nnu}, UNETR~\cite{hatamizadeh2022unetr}, Swin UNETR~\cite{hatamizadeh2021swin}, and Pengy U-Net~\cite{xie2021head}. Even though the plain U-Net was proposed nine years ago, it remains a highly effective tool in the field of medical image segmentation. In the HECKTOR 2021 challenge, almost all participating teams utilized U-Net-based models~\cite{andrearczyk2021overview}. nnU-Net is a UNet-based segmentation method that automatically configures its preprocessing, network architecture, training, and post-processing to suit new tasks. UNETR employs a transformer architecture as its encoder to extract sequence representations from the input volume, allowing it to capture global multi-scale information. It also maintains a "U-shaped" network design that connects encoder and decoder components, similar to traditional U-Net models. Similarly, Swin UNETR applies a swin transformer encoder to extract features at five different resolutions by utilizing shifted windows for computing self-attention. Pengy U-Net was the winning model (team ``Pengy'') of HECKTOR 2021 challenge, whose main innovation is the incorporation of Squeeze-and-excitation normalization (SE Norm)~\cite{iantsen2021squeeze} into a U-Net structure. 

To demonstrate the effectiveness of 3D-based diffusion models, we also conducted an experiment to compare the performances of 3D and 2D DDPM. The network structure utilized in 2D DDPM was set to be the same as that of 3D DDPM, except that 3D convolutional layers and down/up-sample layers were replaced by the corresponding 2D operations. 

In the previous experiments, the input for all models consisted of dual-modality information obtained from both [$^{18}$F]F-FDG PET and CT volumes, aiming to leverage the complementary information available from these distinct modalities. We further carried out an experiment to exam the effect of different modalities on the segmentation task by training the proposed 3D diffusion model using three different sets of input: 3D FDG-PET only, 3D CT only, and 3D FDG PET combined with CT. All other aspects of the training process were kept be the same when varying the network input.

\subsection{{\color{black}Implementation Details}}

{\color{black}
Due to the substantial memory requirements of processing 3D multi-modality volumes, the batch size was set to 1. Moreover, to reduce the memory consumption and accelerate the training speed, we applied random cropping to the original volume, generating small volumes of size $32\times96\times96$ to train the 3D DDPM. Additionally, we employed data augmentation techniques, such as rotation and flipping, to enhance the diversity of the training data. The learning rate was set to $1\times10^{-4}$, and a dropout rate of 0.1 was applied.

To ensure a fair comparison, all 3D models processed small 3D volumes with the size consistent with 3D DDPM. The 2D DDPM was trained using 2D patches with the size of $96\times96$, which were randomly cropped from 2D slices of the transverse view. The number of trainable parameters for each model was set to be roughly equivalent to that of the 3D DDPM. This was achieved by modifying the number of channels in each layer, while keeping the overall structure of the models unchanged. Moreover, training parameters, including optimizer selection and learning rate, were standardized to match those employed in the 3D DDPM. The validation set consisting of 18 cases was utilized to select the optimal checkpoints for all models in our study. The training of all the models was based on a single RTX8000 GPU (48 GB memory), the PyTorch platform and the Adam optimizer. 
}

\subsection{Data Analysis}

We adopted six different metrics to quantitatively evaluate the segmentation results, including the Dice similarity coefficient (DSC), 95th percentile Hausdorff distance (HD95), sensitivity, specificity, false positive rate (FPR), and false negative rate (FNR). DSC is a widely employed metric in segmentation tasks to measure the volumetric overlap between the segmentation output and the ground truth, defined as follows:
\begin{equation}
DSC(\hat{x}_0, x_0^{GT}) = \frac{2 | \hat{x}_0 \cap x_0^{GT} |}{| \hat{x}_0 | + | x_0^{GT} |},
\end{equation}
where $\hat{x}_0$ and $x_0^{GT}$ denote the output and ground-truth segmentation masks, respectively. HD measures the distance between the surfaces of the predicted tumor volumes and the surfaces of the ground-truth tumor volumes. To reduce the impact of outliers, the 95th percentile Hausdorff distance (HD95) is often preferred, calculated as follows:
\begin{align}
HD95(\hat{x}_0, x_0^{GT}) = P_{95}(&\max_{x \in \hat{x}_0} \min_{y \in x_0^{GT}} d(x,y), \max_{y \in x_0^{GT}} \min_{x \in \hat{x}_0} d(x,y)),
\end{align}
where $d(x,y)$ denotes the Euclidean distance between points $x$ and $y$, and $P_{95}$ is an operator extracting the 95th percentile. Furthermore, sensitivity and specificity are also widely utilized metrics. Specificity measures the true negative rate, indicating the model's ability to correctly identify non-tumor regions, while sensitivity measures the model's ability to accurately segment tumor regions. Sensitivity and specificity are calculated as follows:
\begin{align}
Sensitivity &= \frac{TP}{TP + FN}, \\
Specificity &= \frac{TN}{FP + TN},
\end{align}
where $TP$, $TN$, $FP$, and $FN$ represent numbers of true positives, true negatives, false positives, and false negatives, respectively. 
To better illustrate the false classification rates for each method, we also reported the FPR and FNR as follows:
\begin{align}
FPR &= 1 - Specificity = \frac{FP}{FP + TN}, \\
FNR &= 1 - Sensitivity = \frac{FN}{FN + TP}.
\end{align}

Balancing false positive and false negative errors is vital for a reliable and accurate tumor segmentation model. All metrics were computed based on the segmentation results in 3D volumes with size of $144 \times 144 \times 144$.

\section{Results}
\subsection{Comparison with Reference Methods}

\begin{figure*}
	\centering
	\includegraphics[width=.99\textwidth]{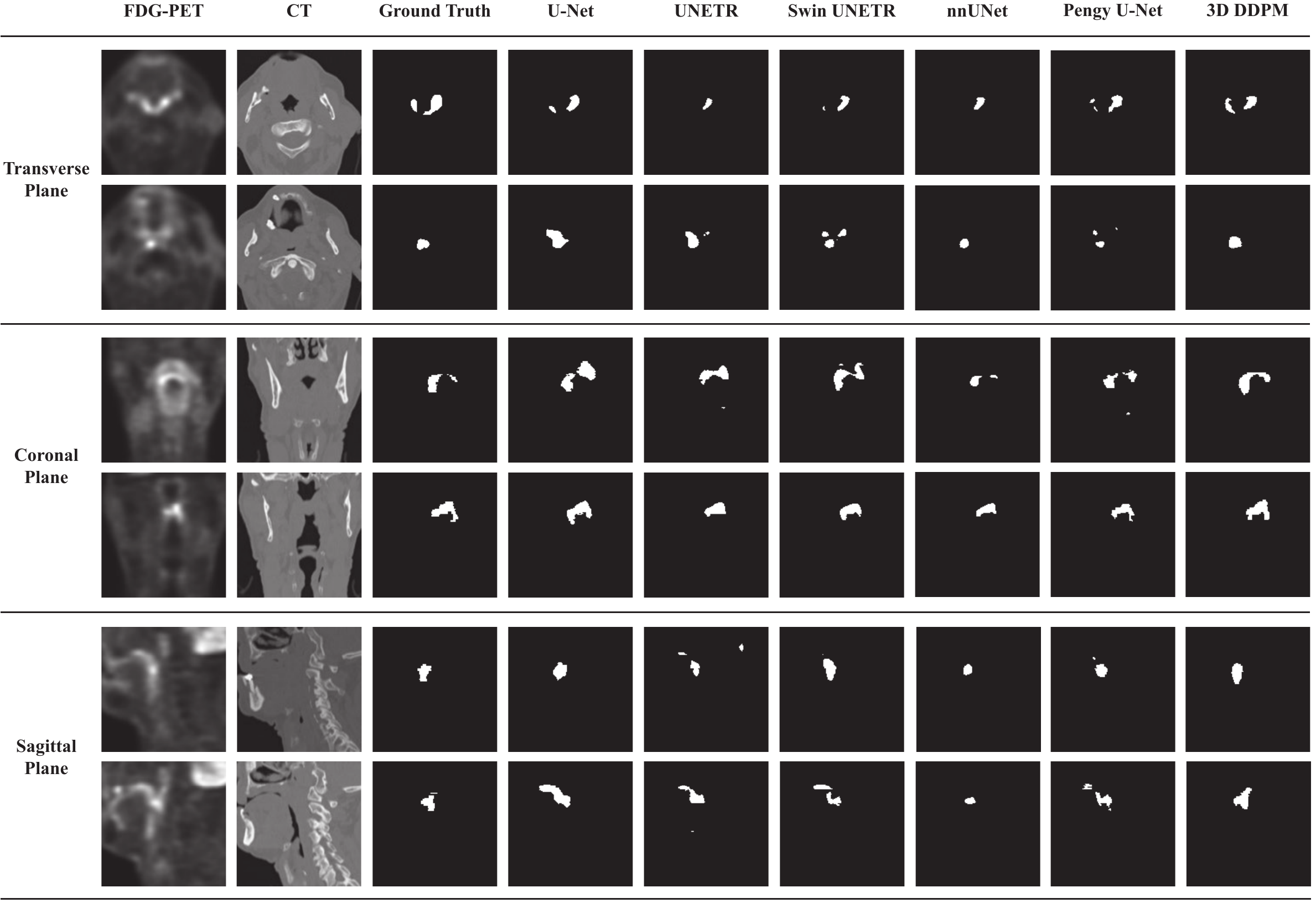}
	\caption {Three views of the H\&N segmentation results generated by different methods.}
	\label{fig_results}
\end{figure*}

\begin{table*}
\centering
\caption{Comparison of metrics from different models (mean $\pm$ standard deviation, median value for HD95). Note that $\uparrow$ indicates that larger values provide better results, while $\downarrow$ indicates the opposite. \textbf{Bold} indicates the best performance, while \underline{underline} indicates the second best.} 
\resizebox{\textwidth}{!}{
\begin{tabular}{@{}lcccccc@{}}
\toprule
          & DSC ($\uparrow$) & HD95 (mm) ($\downarrow$) & Sensitivity   ($\uparrow$) & Specificity   ($\uparrow$)  & FPR ($\downarrow$)  & FNR ($\downarrow$) \\ \midrule
U-Net      & \makecell{0.653\\ ($\pm$0.198)}  & \makecell{54.775}   & \makecell{\textbf{0.765}\\($\pm$0.186)}    & \makecell{0.997\\($\pm$0.003)}    &   \makecell{0.0031\\($\pm$0.0031)}  & \makecell{\textbf{0.235}\\($\pm$0.185)} \\
UNETR     & \makecell{0.684\\($\pm$0.188)}  & \makecell{18.400}  & \makecell{0.701\\($\pm$0.198)}  & \makecell{\underline{0.998}\\($\pm$0.001)}  & \makecell{0.0013\\($\pm$0.0014)} & \makecell{0.299\\($\pm$0.198)} \\
Swin UNETR & \makecell{0.700\\($\pm$0.189)}  & \makecell{8.862}  & \makecell{0.725\\($\pm$0.211)}  & \makecell{\underline{0.998}\\($\pm$0.002)}  &  \makecell{0.0014\\($\pm$0.0017)} & \makecell{0.275\\($\pm$0.212)}\\
nnU-Net    & \makecell{\underline{0.726}\\($\pm$0.188)}  & \makecell{\underline{7.071}}  & \makecell{0.679\\($\pm$0.218)} & \makecell{\textbf{0.999}\\($\pm$0.001)}  & \makecell{\textbf{0.0005}\\($\pm$0.0006)} & \makecell{0.321\\($\pm$0.218)}\\
Pengy U-Net  & \makecell{0.703\\($\pm$0.164)}  & \makecell{18.871}  & \makecell{0.708\\($\pm$0.199)}  & \makecell{\textbf{0.999}\\($\pm$0.001)}  & \makecell{0.0013\\($\pm$0.0015)} & \makecell{0.292\\($\pm$0.199)}\\
3D DDPM      & \makecell{\textbf{0.739}\\($\pm$0.136)}  & \makecell{\textbf{5.000}}  & \makecell{\underline{0.739}\\($\pm$0.188)}  & \makecell{\textbf{0.999}\\ ($\pm$0.001)}  &  \makecell{\underline{0.0011}\\($\pm$0.0014)}& \makecell{\underline{0.260}\\($\pm$0.188)} \\ \bottomrule
\end{tabular}}
\label{tab_fivemodels}
\end{table*}

\begin{figure*}
	\centering
	\includegraphics[width=.65\textwidth]{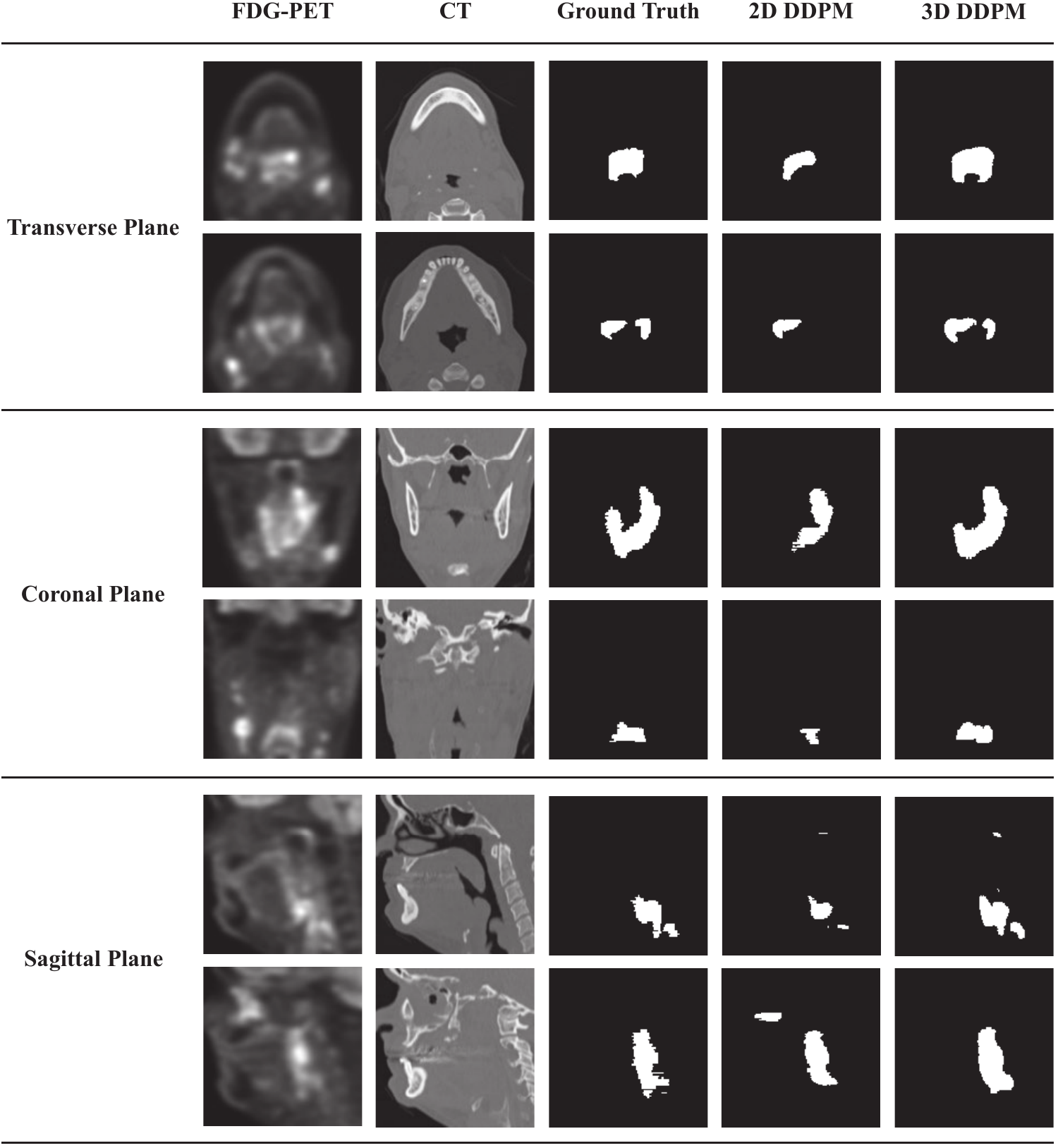}
	\caption {Three views of H\&N segmentation results from the diffusion models in 2D and 3D forms.}
	\label{fig_results_2d}
\end{figure*}

The qualitative and quantitative evaluation results of the proposed method and the reference methods are shown in Figure~\ref{fig_results} and Table~\ref{tab_fivemodels}, respectively. In Table~\ref{tab_fivemodels}, we calculated the average and standard deviation values across all testing cases for each method. For the HD95 metric, we calculated the median value instead due to its upper limit being infinity. The quantitative results showed that the proposed 3D DDPM achieved the highest DSC (0.739) and lowest HD95 (5.000) among all the reference methods; meanwhile, the nnU-Net attained the second-best DSC and HD95. We also noticed that incorporating the Transformer structure in UNETR could improve the performance as compared to the plain U-Net. The Swin UNETR, which employed the advanced Swin transformer structure, could further surpasses the performance of UNETR. Moreover, due to the low proportion of tumor regions in the volumes, all models achieved high specificity. The plain U-Net had the lowest FNR but the highest FPR, indicating that its segmentation results suffer from a high number of false positive errors. Conversely, the nnU-Net achieved the lowest FPR but the highest FNR, indicating a high risk of misclassifying tumors as background. The proposed 3D DDPM model attained the second-best results in sensitivity, FNR and FPR, suggesting a better balance between false positive and false negative errors.

This trend can also be observed in Figure~\ref{fig_results}. Referring to Table~\ref{tab_fivemodels}, the plain U-Net model produced the highest FPR and the lowest FNR. Consequently, as shown in Figure~\ref{fig_results}, the plain U-Net tended to classify more pixels as tumor regions, resulting in more false positive errors. Conversely, the nnU-Net exhibited high FNR, as observed in the qualitative results, leading to the under-segmentation that missed numerous tumor regions. In comparison to these methods, the proposed 3D DDPM model attained the optimal balance between false positive and false negative errors, rendering it more effective for H\&N tumor segmentation.

\begin{table*}
\centering
\caption{Comparison of metrics from the 3D and 2D diffusion models (mean $\pm$ standard deviation, median value for HD95). Note that $\uparrow$ indicates that larger values provide better results, while $\downarrow$ indicates the opposite. \textbf{Bold} indicates the best performance.} 
\resizebox{\textwidth}{!}{
\begin{tabular}{@{}lcccccc@{}}
\toprule
          & DSC ($\uparrow$) & HD95 (mm) ($\downarrow$) & Sensitivity   ($\uparrow$) & Specificity   ($\uparrow$) & FPR ($\downarrow$)  & FNR ($\downarrow$) \\ \midrule
2D DDPM    & \makecell{0.669\\($\pm$0.185)}  & \makecell{12.042}  & \makecell{0.649\\($\pm$0.221)}  & \makecell{\textbf{0.999}\\ ($\pm$0.001)}  &  \makecell{\textbf{0.0011}\\($\pm$0.0012)}& \makecell{0.351\\($\pm$0.221)} \\
3D DDPM      & \makecell{\textbf{0.739}\\($\pm$0.136)}  & \makecell{\textbf{5.000}}  & \makecell{\textbf{0.739}\\($\pm$0.188)}  & \makecell{\textbf{0.999}\\ ($\pm$0.001)}  &  \makecell{\textbf{0.0011}\\($\pm$0.0014)}& \makecell{\textbf{0.260}\\($\pm$0.188)} \\   \bottomrule
\end{tabular}}
\label{tab_2D3D}
\end{table*}

\subsection{Comparison with 2D DDPM}

Quantitative and qualitative comparisons of 2D and 3D DDPM are presented in Table~\ref{tab_2D3D} and Figure~\ref{fig_results_2d}, respectively. From the quantitative results in Table~\ref{tab_2D3D}, 3D DDPM achieved the best performance across all metrics. Qualitative results indicated that 2D DDPM tended to under-segment compared to 3D DDPM, consistent with the high FNR (0.351) of 2D DDPM. These results demonstrate the importance of the information present among neighboring slices for image segmentation tasks, and also the benefits of further extending diffusion models from 2D to 3D.

\subsection{Comparison with Single Modality-Based Segmentations}

\begin{figure*}
	\centering
	\includegraphics[width=.90\textwidth]{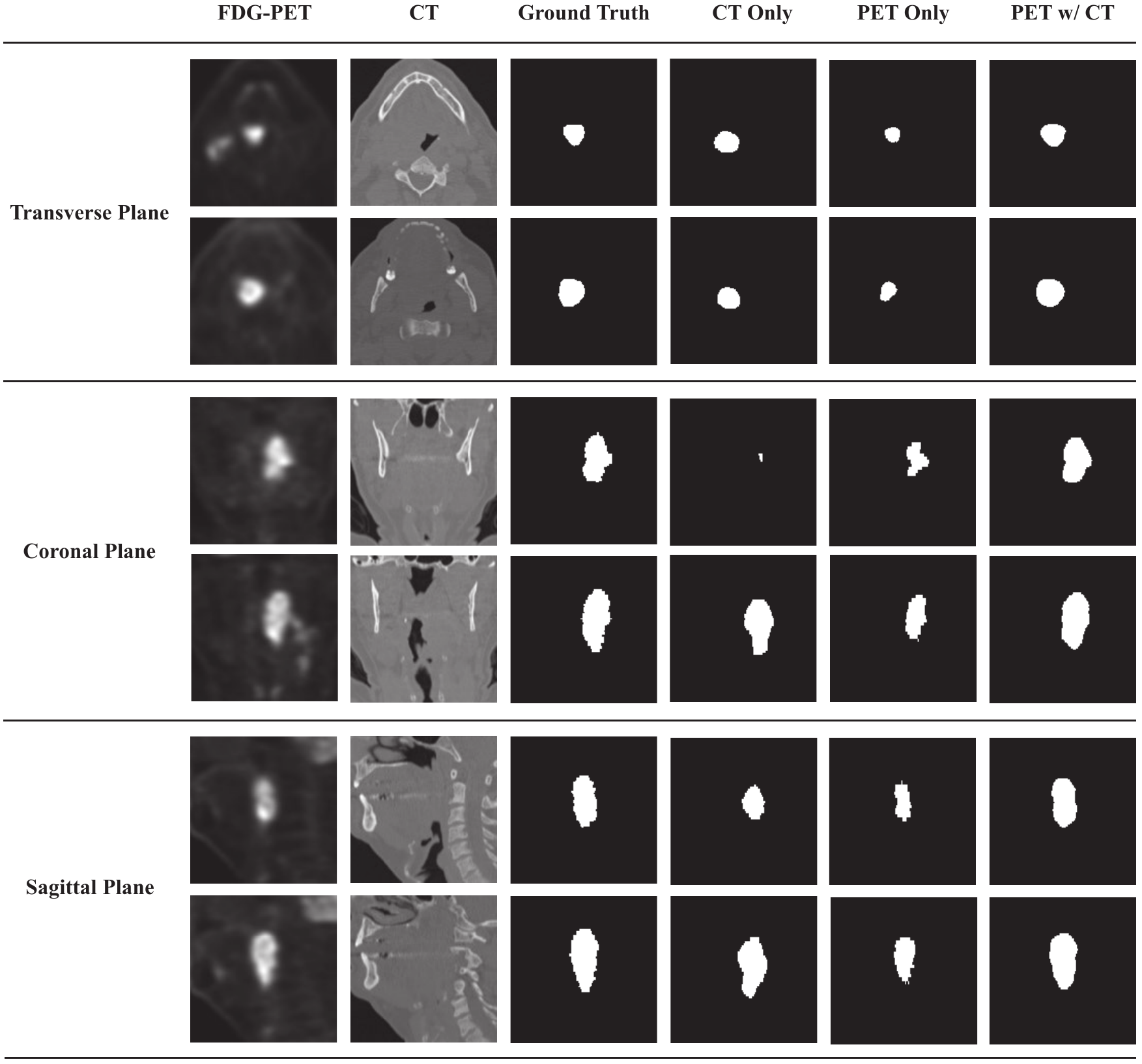}
	\caption {Three views of H\&N segmentation results from the 3D diffusion models with different input modalities.}
	\label{fig_results_petct}
\end{figure*}

Quantitative results of the 3D diffusion model with different network inputs are presented in Table~\ref{tab_petct}. The comparative results clearly highlighted the superior performance achieved with dual-modality input (mean DSC of 0.739 compared to less than 0.570). Furthermore, the model utilizing PET-only input achieved the second-best results, whereas the model relying on CT-only input performed significantly worse compared to the other two. This observation showed that [$^{18}$F]F-FDG PET was more effective at delineating H\&N tumors with high metabolic activities than CT images. These observations were consistent with the qualitative results presented in Figure~\ref{fig_results_petct}. This experiment showed that integrating information from different modalities could effectively enhance the accuracy of H\&N tumors delineation. 

\begin{table*}
\centering
\caption{Comparison of metrics from the 3D diffusion models with different modality inputs (mean $\pm$ standard deviation, median value for HD95). Note that $\uparrow$ indicates that larger values provide better results, while $\downarrow$ indicates the opposite. \textbf{Bold} indicates the best performance, while \underline{underline} indicates the second best.}
\resizebox{\textwidth}{!}{ 
\begin{tabular}{@{}lcccccc@{}}
\toprule
          & DSC ($\uparrow$) & HD95 (mm) ($\downarrow$) & Sensitivity   ($\uparrow$) & Specificity   ($\uparrow$) & FPR ($\downarrow$)  & FNR ($\downarrow$)\\ \midrule
CT Only    & \makecell{0.352\\($\pm$0.224)}  & \makecell{47.021}  & \makecell{0.321\\($\pm$0.208)}  & \makecell{\underline{0.998}\\ ($\pm$0.001)}  &  \makecell{\underline{0.0016}\\($\pm$0.0015)}& \makecell{0.679\\($\pm$0.208)}\\
FDG-PET Only      & \makecell{\underline{0.570}\\($\pm$0.221)}  & \makecell{\underline{24.145}}  & \makecell{\underline{0.638}\\($\pm$0.244)}  & \makecell{0.997\\ ($\pm$0.003)}  &  \makecell{0.0031\\($\pm$0.0034)}& \makecell{\underline{0.362}\\($\pm$0.243)}\\   
FDG-PET w/ CT      & \makecell{\textbf{0.739}\\($\pm$0.136)}  & \makecell{\textbf{5.000}}  & \makecell{\textbf{0.739}\\($\pm$0.188)}  & \makecell{\textbf{0.999}\\ ($\pm$0.001)}  &  \makecell{\textbf{0.0011}\\($\pm$0.0014)}& \makecell{\textbf{0.260}\\($\pm$0.188)} \\ \bottomrule
\end{tabular}}
\label{tab_petct}
\end{table*}

\subsection{Comparison of Uncertainty from DDPM-based Methods}
\begin{figure*}
	\centering
	\includegraphics[width=.90\textwidth]{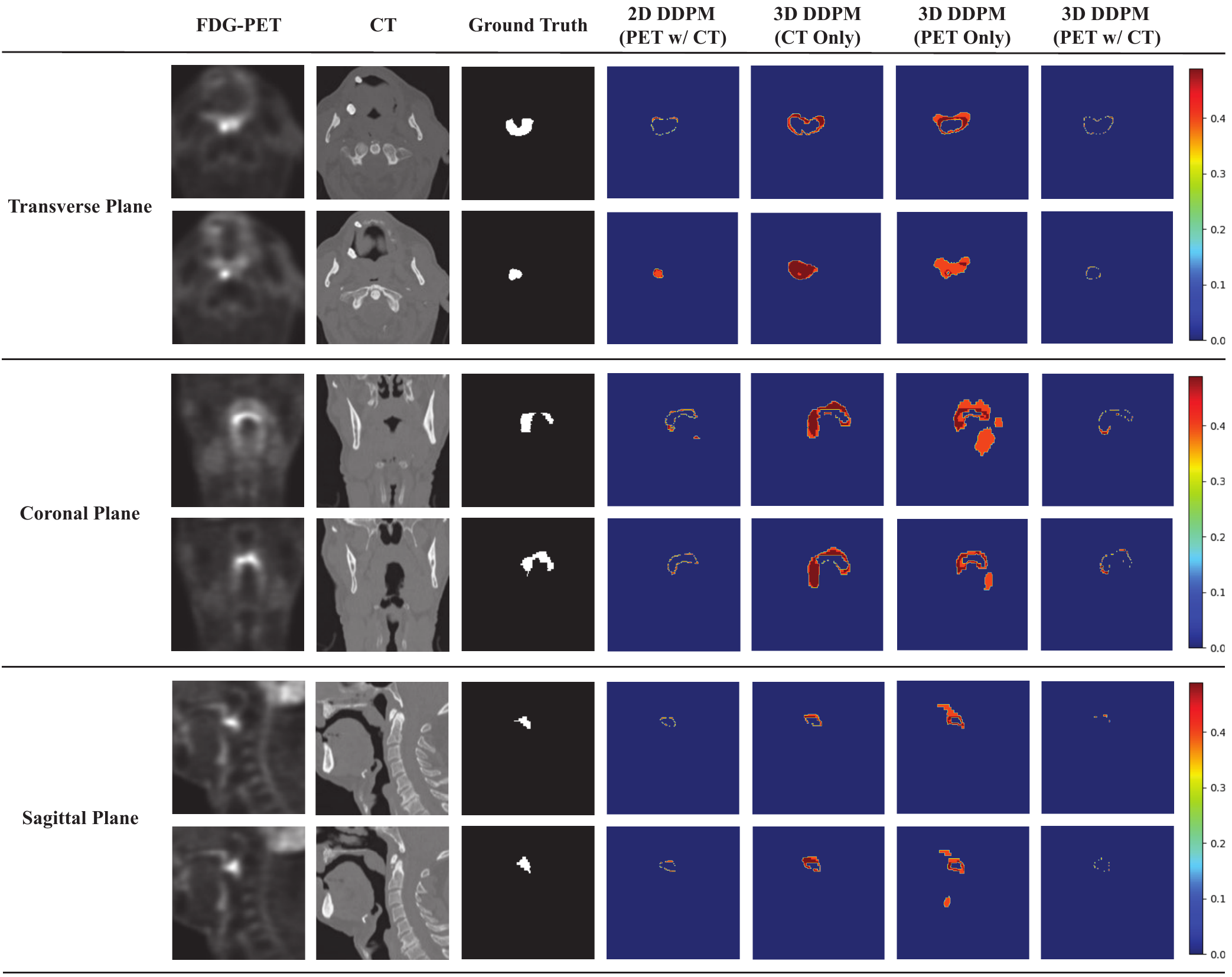}
	\caption {Three views of the uncertainty maps of H\&N tumor segmentation results from different DDPM-based methods.}
	\label{fig_uncertainty}
\end{figure*}

\begin{table}[t]
\centering
\caption{Comparison of uncertainty in H\&N tumor segmentation results from different DDPM-based methods.}
\begin{tabular}{@{}lcccc@{}}
\toprule
           & \makecell{2D DDPM\\(FDG-PET w/ CT)} & \makecell{3D DDPM\\(CT Only)} & \makecell{3D DDPM\\(FDG-PET Only)} & \makecell{3D DDPM\\(FDG-PET w/ CT)} \\ \midrule
Standard Deviation & 0.00041 & 0.00168 & 0.00271 & 0.00024 \\ \bottomrule
\end{tabular}
\label{tab_std}
\end{table}

Stochasticity is one intrinsic property of diffusion models due to the randomly sampled Gaussian noise $x_T$ injected during the inference process. Consequently,  for each inference process,  the generated $x_0$ will have variations, even with the same conditional data. Understanding the degree of uncertainty helps in assessing the stability and consistency of the model's predictions, which is crucial in medical imaging and segmentation tasks. 
{\color{black} To explore the impact of this stochasticity, we repeated the inference process ten times, generating ten different predictions for each case. To represent the uncertainty of each case, we calculated the standard deviation across these ten predictions. We then averaged the standard deviations across all cases to illustrate the overall uncertainty of each method.}
The models involved in this experiment included both 2D and 3D diffusion models with dual-modality inputs, along with 3D diffusion models with single-modality inputs. The qualitative and quantitative results of uncertainty are illustrated in Figure~\ref{fig_uncertainty} and Table~\ref{tab_std}. From the results, it is noteworthy that adding dual-modality inputs could significantly reduce the uncertainty of the diffusion model when comparing 2D/3D DDPM (PET and CT) to  3D DDPM (PET or CT only). The standard deviation of the results from the 3D DDPM (PET and CT) is ten times lower than that of the 3D DDPM (PET only), with values of 0.00024 and 0.00271, respectively. Furthermore, compared with 2D DDPM (PET and CT), it is obvious that 3D DDPM (PET and CT) could further reduce the uncertainty of the generated H\&N  tumor segmentation masks. These observations confirmed that employing 3D operations and dual-modality inputs in diffusion models can effectively reduce stochasticity, resulting in the generation of more robust outcomes.

\section{Discussion}
In this work,  we extended the diffusion model to 3D form for the automatic H\&N tumor segmentation based on 3D PET and CT volumes.  Validations were conducted based on the publicly available HECKTOR 2021 dataset. Quantitative results showed that the proposed 3D diffusion model outperformed other reference models based on the U-Net and Transformer architectures (plain U-Net, UNETR, Swin UNETR, nnU-Net and Pengy U-Net), achieving the best DSC and HD95 values.  Furthermore,  the proposed 3D diffusion model achieved the best balance between false positive and false negative errors. Experiments were also conducted to compare the performances of the diffusion model in 2D and 3D forms, highlighting the effectiveness of extracting 3D features from 3D PET and CT volumes. Performances of the 3D diffusion model with dual-modality and single-modality inputs were also evaluated, which demonstrated the effectiveness of leveraging complementary information from different imaging modalities. Additionally, the comparison of uncertainty provided further evidence of the capability of 3D operation and dual-modality input in effectively reducing the stochasticity.

Thanks to its strong capacity for image generation, the diffusion model has been applied to image segmentation tasks~\cite{wu2022medsegdiff,li2024generic,zhao2024dtan,khosravi2023few}. However, these models operate primarily in 2D space, whereas our experimental results highlight the benefits of 3D operations. While the diffusion model has demonstrated outstanding performance, its primary drawbacks are the high computational demand and slow processing speed. As previously explained, during the reverse process, the clean data $x_0$ was gradually reconstructed from the initial random Gaussian distribution $x_T$ step by step. Typically, the time step $T$ was set to 1000~\cite{ho2020denoising} and thus the reverse process needed to iterate 1000 times during each inference. In our experiment, using one RTX8000 GPU, the 3D diffusion model took approximately 14.7 minutes to complete the inference for one case, regardless of the modality inputs, while the 2D diffusion model took 10.8 minutes. Both models were significantly slower than the U-Net-based models. To address this issue, researchers have proposed various techniques to reduce the number of time steps, e.g. , DDIM~\cite{song2020denoising}, DPMSolver~\cite{lu2022dpm}, and Analytic-DPM~\cite{bao2022analytic}.  In the future, we plan to explore these techniques to reduce the inference time of the diffusion model, making it more efficient and practical for the image segmentation task. 

In this work, the dataset utilized for training and testing all models was the HECKTOR 2021 dataset, which included 224 cases from 5 centers worldwide. In our future work, we plan to train and evaluate the proposed framework based on datasets from additional cohorts to encompass a broader spectrum of patients, which can contribute to more comprehensive and robust model training and validation. Additionally, the HECKTOR 2021 dataset focused on oropharyngeal cancers in the head and neck region of patients. Moving forward, we plan to extend our methods to other medical image segmentation tasks, such as lung cancer, liver tumors, and brain lesions. By broadening the application scope, we aim to evaluate the generalizability and robustness of our methods across various anatomical regions and medical imaging challenges.

The diffusion model is a general framework where various neural network structures can be employed to generate the output from the Gaussian noise and the conditional input. In our work, due to limitations in GPU memory and computational speed, we utilized a 3D U-Net model without any attention mechanism within the diffusion framework. Employing a more powerful neural network architect, such as the Transformer structure and the newly proposed structured state space models~\cite{gu2021efficiently,gu2023mamba}, can potentially enhance the performance of the diffusion model, which is also one of our future research directions.

Uncertainty quantification is a critical aspect of many machine learning applications, especially in fields where reliability and risk assessment are crucial, such as healthcare. Due to their generative nature, diffusion models can produce multiple outputs from the same input, offering a potential approach for uncertainty quantification. Future work will explore the use of diffusion models for uncertainty quantification, particularly in comparison with other stochastic uncertainty quantification methods like Monte Carlo Dropout.

\section{Conclusion}
In this work, we developed a 3D diffusion model to perform automated H\&N tumor segmentation from [$^{18}$F]F-FDG PET and CT images. The diffusion model employed a 3D UNet model and utilized the concatenation of 3D PET, CT, and Gaussian noise volumes as the network inputs. Effectiveness of the proposed diffusion model was validated through the comparison with other state-of-the-art methods based on the U-Net and Transformers. Both qualitative and quantitative results demonstrated that the proposed 3D diffusion model could generate more accurate H\&N tumor segmentation masks compared to the other reference methods.

\section{Acknowledgements}
This work was supported by the National Institutes of Health under grants R21AG067422, R03EB030280, R01AG078250, and P41EB022544.

\section*{References}

\end{document}